\documentclass[10pt,twocolumn,letterpaper]{article}

\usepackage{wacv}              

\usepackage{graphicx}
\usepackage{amsmath}
\usepackage{amssymb}
\usepackage{booktabs}

\usepackage{multirow}
\usepackage{makecell}
\usepackage{graphicx}
\usepackage{bbding}
\usepackage{color}
\usepackage{cite} 
\usepackage[pagebackref,breaklinks,colorlinks]{hyperref}
\usepackage[accsupp]{axessibility} 

\usepackage[capitalize]{cleveref}
\crefname{section}{Sec.}{Secs.}
\Crefname{section}{Section}{Sections}
\Crefname{table}{Table}{Tables}
\crefname{table}{Tab.}{Tabs.}

\def\wacvPaperID{126} 
\def\confName{WACV}
\def\confYear{2024}

\begin{document}

\title{LAVSS: Location-Guided Audio-Visual Spatial Audio Separation
}

\author{Yuxin Ye$^1$, Wenming Yang$^{1}$\thanks{Corresponding author.}, Yapeng Tian$^2$ \\
$^1$Shenzhen International Graduate School, Tsinghua University, China\\
$^2$Department of Computer Science, The University of Texas at Dallas, USA\\
\tt\small yeyx21@mails.tsinghua.edu.cn, yangelwm@163.com, yapeng.tian@utdallas.com
}
\maketitle

\begin{abstract}
Existing machine learning research has achieved promising results in monaural audio-visual separation (MAVS). However, most MAVS methods purely consider what the sound source is, not where it is located. This can be a problem in VR/AR scenarios, where listeners need to be able to distinguish between similar audio sources located in different directions.
To address this limitation, we have generalized MAVS to spatial audio separation and proposed LAVSS: a location-guided audio-visual spatial audio separator. LAVSS is inspired by the correlation between spatial audio and visual location. We introduce the phase difference carried by binaural audio as spatial cues, and we utilize positional representations of sounding objects as additional modality guidance. We also leverage multi-level cross-modal attention to perform visual-positional collaboration with audio features.
In addition, we adopt a pre-trained monaural separator to transfer knowledge from rich mono sounds to boost spatial audio separation. This exploits the correlation between monaural and binaural channels.
Experiments on the FAIR-Play dataset demonstrate the superiority of the proposed LAVSS over existing benchmarks of audio-visual separation. Our project page: \textcolor{red}{\url{https://yyx666660.github.io/LAVSS/}}.
\end{abstract}

\section{Introduction}

Auditory and visual characteristics can convey important semantic and spatial information, which plays a crucial role in audio-visual separation \cite{survey}. The well-known cocktail party problem\cite{cocktail} is a classical task of sound source separation \cite{watchingunlabel,sop,ccol} and localization \cite{Beamforming, 3D}. It aims at separating the target source audio from the given audio mixture. A popular line of work for audio-visual separation is to encode visual information as guidance for resolving sound ambiguity from mixed audio sources \cite{cof,mp-net,sep-fusion}.
\begin{figure}[bthp]
\begin{minipage}[b]{1.0\linewidth}
  \centering
  \centerline{  \includegraphics[width=7.5cm]{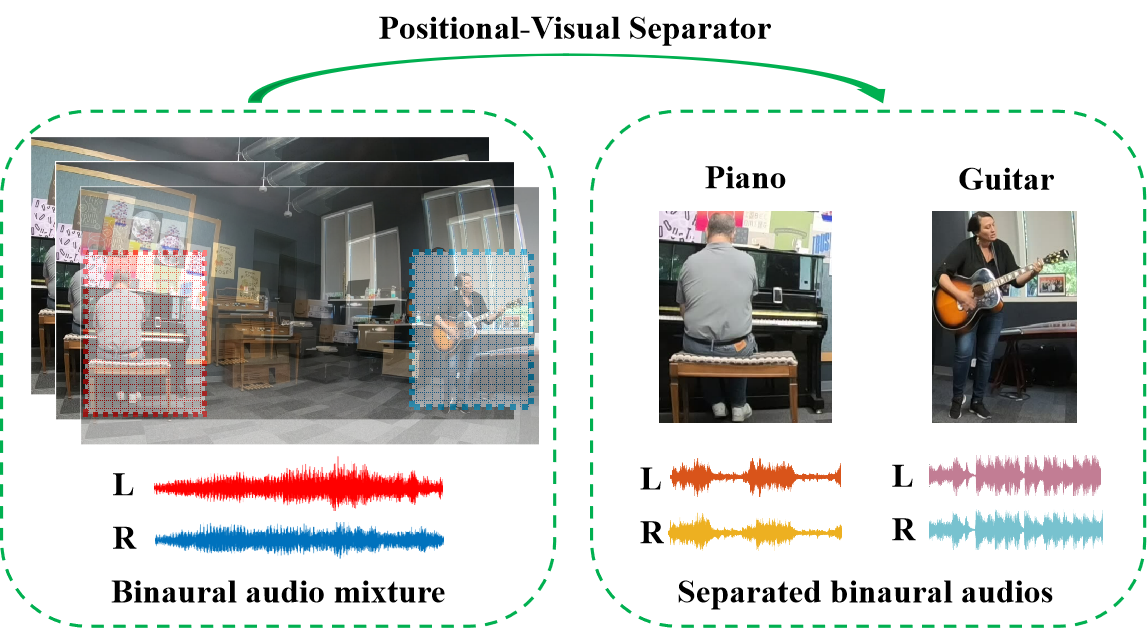}}
\end{minipage}
\caption{Our LAVSS can separate individual binaural sounds for sounding objects (piano and guitar) from a binaural audio mixture.}
  \vspace{-3mm}
\label{teaser fig}
\end{figure}
For instance, lip motion \cite{visualvoice, listenlook} and facial expression \cite{lookingintospeech} information were applied to separate speech sounds from different speakers. Motion \cite{motion, visuallymotion} and gesture \cite{gesture} appearance features were exploited to guide music sound separation. Other methods utilize instrument category \cite{co-sep,graph} or multimodal attention \cite{audioscopev2, audioscope, iquery} to leverage the association between visual and audio modalities.

Predominant audio-visual separation (AVS) methods have typically been designed for monaural audio-visual separation (MAVS). However, scenarios limited to single-channel audio lack the capacity for perceiving 3D visual scenes accompanied by spatial audio. Although being attempted earlier in \cite{2.5D}, researches on \textit{audio-visual spatial audio separation} (AVSS) (see \cref{teaser fig}) are highly limited. For example, audiences can discern the orientation of the piano and guitar since they hear the mixed spatial audio with varying acoustic intensities for each ear \cite{2.5D}. Unlike MAVS, AVSS provides listeners with a more immersive perceptual experience, thus making it a novel and challenging task.

Existing spatial audio-visual works have mainly focused on spatial audio generation \cite{2.5D,genewithout,consist-gen,360}. This involves converting standard monaural audio into binaural or ambisonic sounds. Sep-stereo \cite{sep-stereo} regards MAVS as a specific case of binaural audio reconstruction at the cost of artificially rearranging visual information. 
However, these methods lack sufficient audio-visual modeling and still exhibit a domain gap when it comes to spatial audio separation. 

In this paper, we address the audio-visual spatial audio separation task by simultaneously considering \textbf{what and where} the sounding object is. In an effort to overcome current limitations, we introduce a new Location-Guided Audio-Visual Spatial Audio Separation (LAVSS) method. We first detect sounding objects to obtain regional visual embeddings (what). Then we encode the spatial location of the sounding objects explicitly. The positional embeddings can be another guidance to reveal the spatial information (where), which benefits separating individual audio in different directions. 
How does this correspond to audio? Since binaural audio carries spatial information cues, we consider the inter-microphone phase difference (IPD) \cite{IPD1,IPD2}, which is commonly used in multi-microphone speech segregation and separation \cite{far-field,WPE-BSS,Speech-segregation}. 
The IPD information represents the established spatial feature between the left and right channel. We force the network to learn the synchronization and correlation between the spectra-spatial audio feature and the visual-positional representations.
Moreover, we propose a multi-scale attention-based fusion network to integrate the visual, positional, and audio features. All constituent modalities work in concert to benefit AVSS. 


Additionally, to leverage the correlation between monaural and binaural channels, we employ a pre-trained separator. This aids in the knowledge transfer from rich mono sounds, thereby enhancing spatial audio separation. By utilizing the extensive video data with monaural sounds available in the MUSIC-21 dataset, we accomplish effective pre-training. Experiments on the binaural FAIR-Play dataset can validate the efficacy of LAVSS. It achieves state-of-the-art performance, particularly in scenarios where similar acoustic sources are positioned in different directions.

\begin{figure*}[htbp]
\begin{minipage}[b]{1.0\linewidth}
  \centering
  \centerline{\includegraphics[width=17cm]{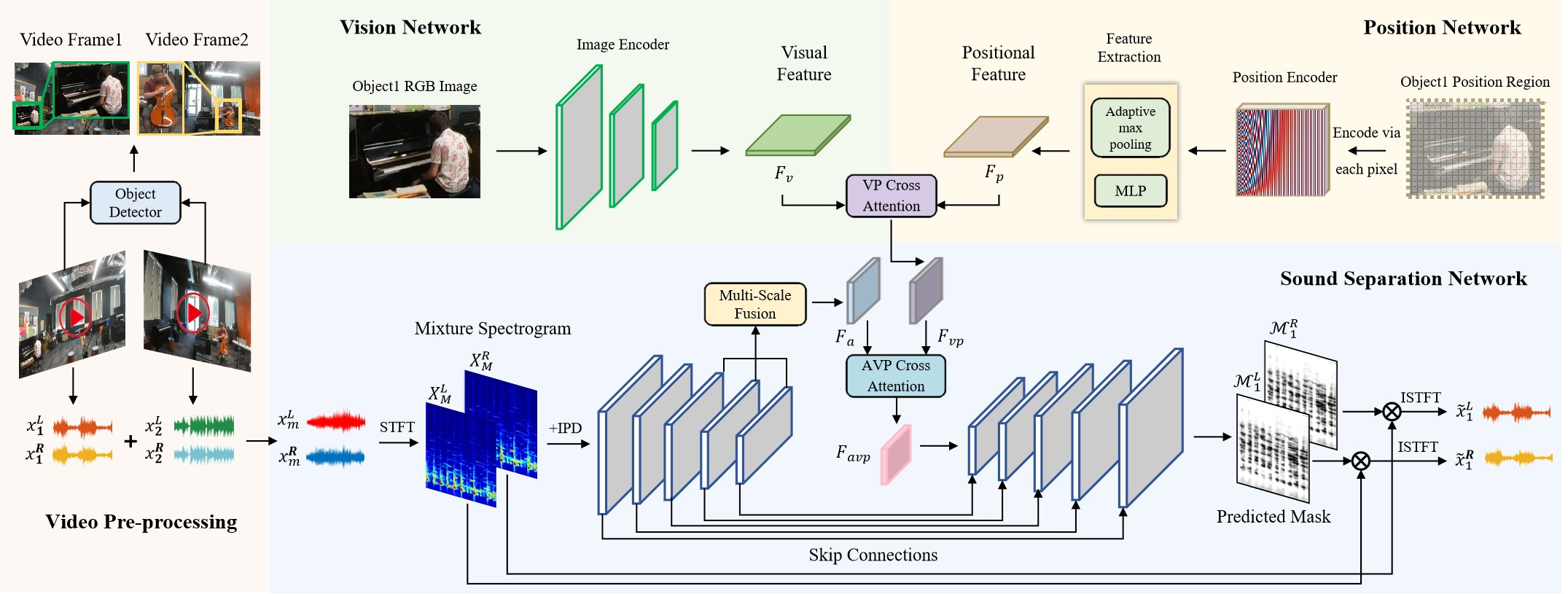}}
\end{minipage}
\caption{An overview of the proposed architecture. Video pre-processing includes object detection and source mixing. Vision extraction network encodes the visual regions of detected objects; position network simultaneously encodes regional coordination features; VP Cross Attention module aggregates visual and positional representations; sound separation network exploits the fused feature as guidance to separate binaural sounds. Note that all the operations are depicted for only one video (piano), the other remains the same during training.}
\label{network}
\end{figure*}

Our contributions are as follows:
i) We put forward a multi-modal framework to address the AVSS task. 
ii)
We take advantage of the correlation between the IPD and positional features, which respectively represent the spatial properties of binaural audio and the explicit location cues of the sounding objects.
iii) We pre-train the separator on an external mono dataset to facilitate AVSS network learning by leveraging the correlation between monaural and binaural channels. 
iv) Experiments demonstrate the superior improvement and generalizability of our LAVSS over state-of-the-art audio-visual separation approaches.

\section{Related Work}
\noindent\textbf{Audio-Visual Learning} \quad
Audio-visual learning has gained considerable interest in recent years, with researchers achieving promising results in a variety of fields. These include self-supervised learning \cite{attend,trackstereo,Ambient-supervise}, audio-visual speech recognition \cite{listenlook,EPIC-Fusion,SCSampler,voice-face,read-cocktail}, visually guided spatial audio generation \cite{2.5D,sep-stereo,consist-gen,genewithout}, audio-visual speech and music separation \cite{watchingunlabel,sop,lookingintospeech,righttotalk} and localization \cite{tian-local,dual-localization,localinvisualscece,Discriminative-local,coarse-local}, as well as environment acoustics learning \cite{AV-NeRF,few-shot,LearningAcoustic}. Unlike these prior works, we make the first attempt to tackle audio-visual spatial audio separation by incorporating visual positional features as an additional modality and employing the cross-modal attention. 

~\\
\noindent\textbf{Audio-Visual Source Separation} \quad
Sound source separation is a crucial part of speech front-end research and music processing. Traditional signal processing methods usually exploit filtering to strengthen source separation \cite{Monaural,2001One,Markov,Nonnegative,wiener,lowrank} and localization \cite{Beamforming, 3D}. 
Machine learning methods like end-to-end speech separation \cite{tasnet,convtesnet,multi} aim at performing waveform transformation in the time domain. The well-known cocktail party problem \cite{cocktail} is a classical task of sound source separation \cite{watchingunlabel,sop,ccol}. Recently the self-supervised visually guided audio-visual source separation has obtained significant attention \cite{sop,co-sep,cof,mp-net,ccol,sep-fusion}. 
For one aspect, most works exploit appearance features as visual guidance. From the whole image frame \cite{sop,watchingunlabel} to detected sounding object regions \cite{co-sep,ccol}, these works focus on how to obtain precise visual features. Other visual appearances such as motion \cite{motion}, gestures \cite{gesture,TriBERT} are exploited to capture the body movement postures of players. Recent works regard the human and instruments as nodes to build the graph relationships between them \cite{graph,graph2}. 
For another aspect, some researchers optimize the architecture of the separation network \cite{mp-net,cof,ccol} and try to fuse visual-audio modality in an effective manner \cite{sep-fusion}. 
For recent studies, vision transformers \cite{iquery,TriBERT,visuallymotion,VoViT} and attentions \cite{audioscope,audioscopev2,attend} are widely used in multi-modal collaboration. From 2D to 3D, active sound separation \cite{Move2Hear,active-sound} for AR/VR scenarios has become promising future research. 
However, methods basically conducted for mono audios have limited capabilities with spatial ones in real scenarios. Different from MAVS approaches, we propose to relate spatial cues of audio and sounding objects to resolve AVSS. 

~\\
\noindent\textbf{Audio-Visual Spatial Audio Generation} \quad
Audio-visual cross-modality generation aims to generate audio from visual signals \cite{2.5D,consist-gen,audiogen,VisualtoSound,Foley,trackstereo,mutualgen,attention-spatialization}. For instance, Zhou \emph{et al.} \cite{VisualtoSound} utilize the synchronization of visual cues and encoders to generate natural sound for videos in the wild. Zhou \emph{et al.} \cite{trackstereo} and Gao \emph{et al.} \cite{2.5D} adopt a U-Net to encode monaural input and decode binaural counterpart through visual guidance at the bottleneck. Sep-stereo \cite{sep-stereo} put forward an associative pyramid structure to better fuse audio and visual modalities for generation stereo. Other methods \cite{audiogen,Foley} generates audio samples conditioned on text inputs, motion key points, and position information \cite{depth_gen, geometry_gen}, respectively.  
Other works concentrate on 360° audio generation and spatialization. Scene-aware audio \cite{sceneaware} can be converted from a single-channel microphone and transformed into spatial audio. Morgado \emph{et al.} \cite{360} take real spatial audio as self-supervision for ambisonic audio generation. 
Different from these works, our main focus lies in spatial audio separation. 

\section{Proposed Method}
\label{sec:guidelines}
\subsection{Overview}
\label{2.1}
Given an unlabeled video segment V and its corresponding spatial audios $x^L(t)$ and $x^R(t)$, the detected audible objects are defined as $\mathcal{O}=\left\{O_1,...,O_N\right\}$ for each video frame. Our spatial audio separation task aims to separate the individual audio of each sounding object from the mixed audio:
$x^L(t)=\sum_{n=1}^N x_{n}^L(t), x^R(t)=\sum_{n=1}^N x_{n}^R(t)$,
where $x_{n}^L(t)$ and $x_{n}^R(t)$ represent the time signals received at both ears of corresponding object sources. 

As depicted in \cref{network}, our LAVSS training architecture consists of four parts: the video pre-processing module, a vision and position network, and a multi-modal sound separation backbone.
During video pre-processing, we utilize two sets of solo videos and their synchronized spatial audios $\left\{V_{1}, x_{1}(t)\right\}$, $\left\{V_{2}, x_{2}(t)\right\}$ with sounding objects $O_{1}, O_{2}$ in both videos \cite{ccol}, we artificially mix two binaural sounds:    $x_{m}^L(t)=x_{1}^L(t)+x_{2}^L(t), x_{m}^R(t)=x_{1}^R(t)+x_{2}^R(t). $ 
Then we perform object detection to obtain the object bounding boxes and the corresponding coordinates of the objects. The vision network encodes the detected objects to produce visual features. For the position network, we conduct positional encoding for each pixel in the visual object region. The visual and positional features represent the semantic and spatial information of the sounding object, respectively. Both features are mapped into a common embedding space and performed attention-based fusion. 
 
The binaural audio mixture is transformed into the time-frequency domain and passed to an encoder-decoder sound separation network, which is pre-trained on an external monaural dataset. We creatively introduce the inherent IPD between the left and right channels for spatial audio separation. The IPD feature and magnitude spectra are concatenated to leverage both spatial and spectral cues of audio, which correspond to the visual and positional features of the object. All features are fused through a multi-scale attention-based fusion module and transformed into time-discrete space. Finally, we obtain the estimated binaural audios $\hat{x}_{n}^L(t), \hat{x}_{n}^R(t)$ of individual objects. More details of our LAVSS are provided in the supplementary material.

\subsection{Vision-Position Embedding Framework}
\noindent \textbf{Vision network}
\quad
In order to precisely localize the audible objects, we choose the widely used detector Faster R-CNN \cite{c1} trained on labeled Open Images dataset \cite{Openimages} used in \cite{co-sep, ccol}. All potential objects $\mathcal{P}=\left\{P_1,...,P_N\right\}$ for each video are detected. 
Given a video frame $V$, detections of all objects consist of four items $\left\{\left(M_{V}^{n}, C_{V}^{n}, P_{V}^{n}, B_{V}^{n}\right)\right\}_{n=1}^{N}=$ FRCNN $(V)$, which represent the frame index $M$,  instrument category $C \in \mathcal{C}$, detection confidence probability $P$ and bounding box $B$ for each detected object. Then we screen out one object with the highest confidence score among all detected ones as the audible object for each solo video frame (top two for duet video). 

\begin{figure*}[htbp]
    \centering
    \begin{subfigure}{0.45\linewidth}
        \centering
        \includegraphics[width=6.2cm]{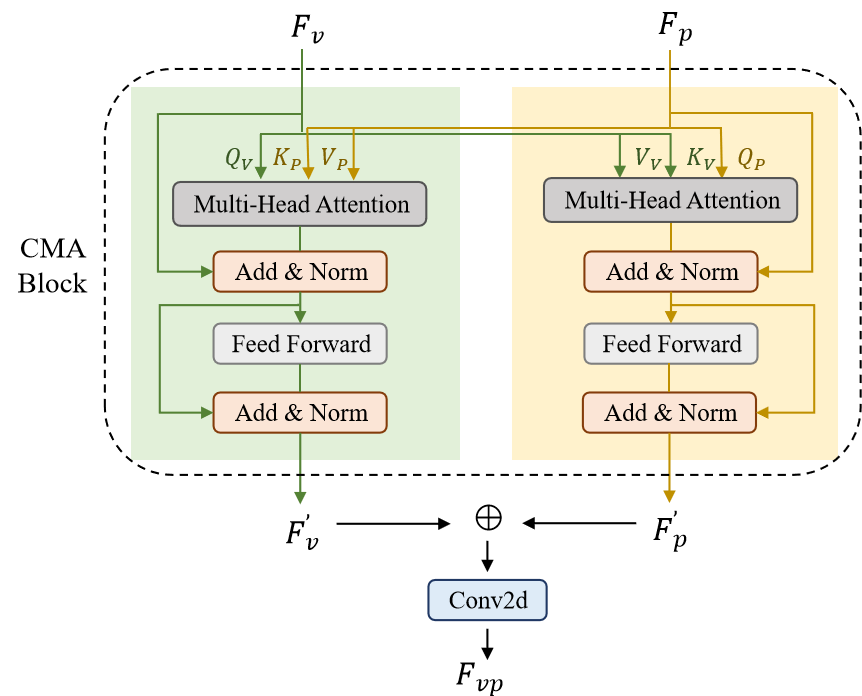}
		\caption{VP Cross-attention module}
		\label{VP_atten} 
    \end{subfigure}
    \begin{subfigure}{0.5\linewidth}
        \centering
        \includegraphics[width=9.2cm]{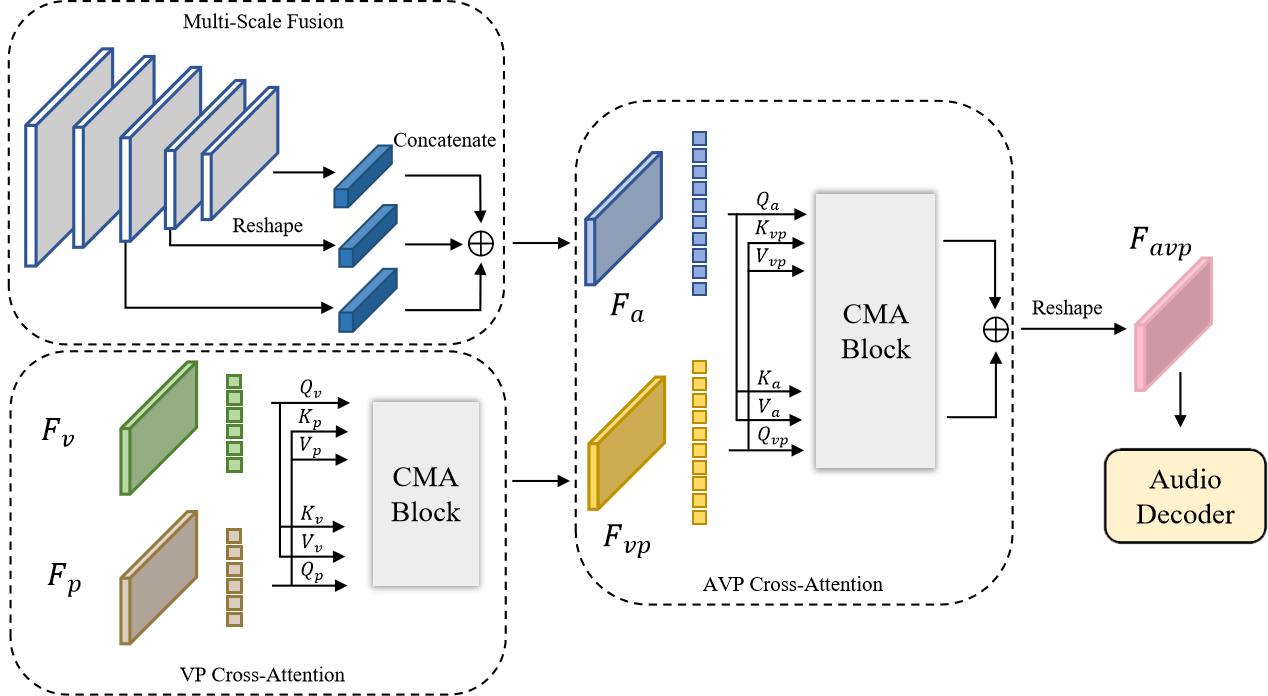}
		\caption{Multi-Scale Audio Fusion Network}
		\label{AVP_atten} 
    \end{subfigure}
	\caption{Two basic blocks of multi-attention modules. (a) VP Cross-Attention, in which the vectors of visual and positional features are integrated through Cross-Modal Attention (CMA) block; (b) The architecture of multi-scale audio fusion network, which consists of the multi-scale fusion, VP/AVP cross-attention modules to introduce the interactions between vision, position and audio modalities.}
	\label{attention}
\end{figure*}

The visual image region for the selected object is of size $3\times H_b\times W_b$, where $H_b, W_b$ denote the height and width of the detected bounding box. For visual feature extraction, objects are resized and passed to a pre-trained ResNet-18 \cite{c8} network. We obtain the visual embedding $F_v\in \mathbb{R}^{C_v\times H \times W}$ before the last fully-connected layer, where $H_b^{'}, W_b^{'}$ represent the resized image shape. $H = H_b^{'}/32, W = W_b^{'}/32, C_v = 512$ denote the feature map size and channel dimension of $F_v$, respectively.

\noindent\textbf{Position network}
\quad
Going beyond the general MAVS strategy, one of the critical innovations of our method is specializing in spatial audio separation. Specifically, we leverage positional representations as a new constituent modality and demonstrate the association with spatial distribution embedded in spatial audio.
Inspired by the positional encoding in Transformer \cite{c6} and NeRF \cite{c7}, we consider how to encode positional representations of audible object regions into a higher dimensional space. 
We leverage a 2D positional encoding for spatial coordinates of detected objects, thus forcing our positional network to approximate a higher frequency function and guide spatial audio separation. Here the function $\gamma(\cdot)$ represents a mapping function from low-dimensional space into a higher one, 
\begin{equation}
\label{posen}
\begin{small}
\begin{aligned}
     \gamma(x, y)=\left(\sin (2^{0}\pi x), \cos (2^{0}\pi x), \sin (2^{0}\pi y), \cos (2^{0}\pi y), \ldots , \right.\\
     \sin (2^{D-1} \pi x), \cos (2^{D-1} \pi x), \sin (2^{D-1} \pi y), \cos (2^{D-1} \pi y))
\end{aligned}
\end{small}
\end{equation}

\noindent This sinusoidal function is applied simultaneously to 2D coordination in $\mathbf{(x, y)}$ (which are normalized to range $[-1,1]$ \cite{c7}) for expanding to higher dimensions via gamma encoding. 
In our experiments, we set $D=16$ for $\gamma(x, y)$ to encode each pixel in the detected object region relative to the whole video frames of size $1280 \times 720$. Then we obtain a tensor of size $C_e\times H_b \times W_b$ by \cref{posen}, where $C_e = 64$ denotes the dimension of the encoded positional embedding. For position feature extraction, the encoded features are performed adaptive max pooling followed by multi-layer perception (MLP). Finally, the positional feature is converted to $F_p \in \mathbb{R}^{C_p\times H\times W}$, where $C_p$ is equal to the vision feature dimension $C_v$ in the previous section.

~\\
\textbf{VP Cross Attention Module}
\quad
For multi-modal modeling, the VP cross-attention module is implemented to integrate the visual and spatial position embeddings. As illustrated in \cref{attention} (a), the VP Cross-Attention module is composed of a CMA block and a convolutional layer. For instance, given an input query $M \in \mathbb{R}^{H_m \times W_m \times D}$ and $N \in \mathbb{R}^{H_n \times W_n \times D}$, $CMA(M, N, N)$ performs cross-modal attention over the first and second axes of $N$, yielding an output tensor of shape $H_m \times W_m \times D$,
\begin{equation}
\begin{aligned}
     \alpha = LN(MHA&(M_Q, N_K, N_V) + M) \\
     \quad CMA(M, N, N) &= LN(FFN(\alpha) + \alpha)
\end{aligned}
\label{CMA}
\end{equation}
\noindent where $M_Q$ is the query vector of $M$, $N_K$, $N_V$ are key and value vectors of $N$. $MHA$, $FFN$, $LN$ denote the multi-head attention, feed-forward layer, and layer normalization, respectively. The $F_v$ and $F_p$ are first passed to the CMA block. Then the visual-positional feature $F_{vp} \in \mathbb{R}^{C_{vp}\times H\times W}$ can be obtained after a convolutional layer to halve the channel dimension. The core part of the module is given by, 
\begin{equation}
\begin{small}
\begin{aligned}
     F_{vp} = Conv(CMA(F_v, F_p, F_p) \oplus CMA(F_p, F_v, F_v))
\end{aligned}
\end{small}
\end{equation}
\noindent where $\oplus$ and Conv denote the concatenate operation and point-wise convolution, respectively.

\subsection{Multi-modal Sound Source Separation}
\noindent \textbf{Audio Embedding Network}
\quad
We follow the widely used mix-and-separate \cite{sop} method and manually mix two video sounds. The time-discrete binaural audio waveform $x_m^L(t), x_m^R(t)$ are first converted to time-frequency spectrograms $X_m^L, X_m^R$ through STFT \cite{c3} transform. Several previous MAVS works \cite{sop, co-sep, ccol} take only log power spectra as the input of the network. In terms of spatial audio, sound source locations are determined by time differences between the sound sources reaching each ear \cite{2.5D, ITDILD}, which can be measured by the inter-microphone phase difference (IPD) between the left and right channels. IPD increases the feature discrimination of location information and indicates the spatial acoustic characteristics of the room. Alternatively, it reveals the different directions of the same sounding objects. The IPD can be calculated as follows,
\begin{equation}
    IPD = cos(\angle X_m^L - \angle X_m^R)
\end{equation}

\noindent where $\angle$ represents the phase angle of the complex spectrogram. One way of utilizing such multi-channel inputs is to feed the network with both log power spectra and IPD features \cite{far-field}. We concatenate both features and obtain the audio embedding of size $2\times T\times F$ for each channel, where T and F represent the time and frequency dimensions, respectively. In this manner, the input of the sound separation network contains both the acoustic spectra (what) and spatial cues (where) carried by the binaural audio. 

Then a U-Net \cite{c9} backbone is used for encoding the composed spectrograms and IPD feature into semantic representations. The architecture is composed of $N$ down- and up-convolutional layers followed by a BatchNorm layer and Leaky ReLU. At the bottleneck, the multi-scale audio fusion network performs multi-modal modeling over the audio, vision, and position features. Note that the sound separation network parameters are shared across left and right channels during training and testing.  

~\\
\textbf{Multi-Scale Audio Fusion Network}
\quad
To establish the relationship between the spectra-spatial audio feature and the visual-positional representations, we put forward a multi-scale audio fusion network visualized in \cref{attention} (b). 
For multi-scale feature fusion, three feature tensors $F_a^{N-i} (N=7, i=0,1,2$) extracted by the last three down-sample convolutional layers are reshaped to $C_a\times Q_a^{N-i}$ by multiplying the time and frequency dimension. Then $f_{Concat}(\cdot)$ performs concatenation along the query dimension to generate audio queries $F_a \in \mathbb{R}^{C_a \times Q_a}$,
\begin{equation}
     F_{a} = f_{Concat}(F_a^N, F_a^{N-1}, ..., F_a^{N-i}), i=0,1,2
\end{equation}

\noindent The audio feature $F_a$ is fed into the AVP cross-attention module to adaptively interact with the visual-positional feature $F_{vp}$. The output audio embedding $F_{avp} \in \mathbb{R}^{C_a\times \frac{T}{S} \times \frac{F}{S}}$ (S denotes stride of audio feature map) is computed by
\begin{equation}
\begin{small}
\begin{aligned}
     F_{avp} = f_2(CMA(F_a, F_{vp}, F_{vp}) \oplus f_1(CMA(F_{vp}, F_a, F_a)))
\end{aligned}
\end{small}
\end{equation}
\noindent where $\oplus$ is the concatenate operation, $f_1(\cdot)$ denotes the one-dimensional convolusion, $f_2(\cdot)$ means dimensional expansion and two-dimensional convolution operation. 
The feature vector $F_{avp}$ is regarded as guidance for audio separation and passed to the decoder up-sample layers of U-Net. Finally, we obtain the predicted magnitude binary masks $\hat{\mathcal{M}_n^L}, \hat{\mathcal{M}_n^R}$, which are multiplied by the original mixture spectrogram $X_m^L, X_m^R$ to produce the final estimation of output spectrograms. The estimated audios $\hat{x}_{n}^L(t), \hat{x}_{n}^R(t)$ are obtained after ISTFT. More specifically,
\begin{equation}
\begin{aligned}
    \hat{x}_{n}^B(t) &= ISTFT(\hat{\mathcal{M}_n^B} \odot X_m^B) \\
    \mathcal{M}_{gt,n}^B(u, v) &= [X_n^B(u, v) \geq X_m^B(u, v)]
\end{aligned}
\end{equation}

\noindent where $\odot$ denotes element-wise multiplication, $(u, v)$ represents time-frequency dimension, $B\in[L, R], n\in[1, 2]$ (number of the objects). The ground truth of binary masks $\mathcal{M}_{gt,n}^B$ are created by the ratio between the source spectrograms $X_n^B$ and the mixture spectrograms $X_m^B$. 

~\\
\textbf{Overall learning Objective}
\quad
We optimize our LAVSS framework training objective by jointly minimizing a combination of both frequency and time reconstruction losses. For the frequency domain loss, we measure the linear combination between the L1 and L2 losses over the predicted ratio masks and ground-truth in \cref{loss}. Furthermore, we introduce the loss between the target audio $x_{n}^B(t)$ and reconstructed audio $\hat{x}_{n}^B(t)$ over the time domain. Formally,
\vspace{-2mm}
\begin{equation}
\begin{small}
\label{loss}
\begin{aligned}
     \mathcal{L}_{freq} =\sum_{n=1}^N\sum_{B\in{L,R}}&\Vert \hat{\mathcal{M}}_{n}^B-\mathcal{M}_{n}^B\Vert_1 
     +\alpha\Vert \hat{\mathcal{M}}_{n}^B-\mathcal{M}_{n}^B\Vert_2   \\
     \mathcal{L}_{time} &= \sum_{n=1}^N\sum_{B\in{L,R}}\Vert \hat{x}_{n}^B(t)-x_{n}^B(t)\Vert_1  \\
     \mathcal{L}_{binaural} &= \mathcal{L}_{freq} + \beta \mathcal{L}_{time} 
\end{aligned}
\end{small}
\end{equation}

\subsection{Transfer learning by external monaural dataset}
Due to the complexity of the binaural attributes, the framework designed for spatial audio is complicated for training directly. To alleviate this issue, we choose a widely used mono dataset MIT MUSIC to perform transfer learning for two reasons. First, the binaural FAIR-Play dataset contains much scarce training data due to the recording difficulty. In contrast, the MUSIC dataset includes more instrument categories and videos, which can mitigate the difficulty of AVSS and make the training more robust. Some of the instrument types overlap, which makes the sound separation between similar acoustic characteristics mutually beneficial. Second, considering the relationship between mono and binaural audio, we can transfer knowledge from rich mono sounds to boost spatial audio separation performance. Thus, a pre-trained monaural separator is adopted by training on the MAVS network backbone in \cite{2.5D}. 

Similar to the training process in \cref{network}, we pre-process the videos in MUSIC dataset. Then we take the monaural mixtures and detected RGB image regions into the U-Net separation and visual network, respectively. Both features are fused by multi-scale attention-based fusion at the bottleneck. Note that the monaural audios do not possess the spatial location information. The IPD and position feature will \textbf{not} be considered as input to the network. After training, the separation network can be a good separator for most mixture audios of different instruments, which simultaneously alleviates network learning for training binaural audios. Finally, we load pre-trained parameters both of the U-Net separation and visual network as initial weights and perform complete position-guided audio-visual separation network training on the FAIR-Play dataset. More details of pre-training are revealed in the supplementary material.

\section{Experiment and Results}

\subsection{Experimental Settings} 
\noindent \textbf{Datasets}
\quad
In our experiments, both monaural and spatial datasets are used for training.
To perform monaural separator pre-training, we use MUSIC dataset~\cite{sop}, which is a commonly used dataset for MAVS. 
It contains 685 solo and duet videos with 11 instrument categories: accordion, acoustic guitar, cello, clarinet, erhu, flute, saxophone, trumpet, tuba, violin, and xylophone. We utilize 520 mono solo videos and split them into train/val/test sets with 468/26/26 for pre-training.  

For spatial audio separation, we use the FAIR-Play dataset~\cite{2.5D}. The instrument type contains cello, guitar, drum, ukelele, harp, piano, trumpet, upright bass, and banjo.
We use 1039 10s solo videos with spatial audio during training and testing. To evaluate the proposed LAVSS model conditioned on the detected object coordination, we randomly split it into train/val/test sets: 728/103/208. Moreover, we evaluate the separation ability of LAVSS for separating multiple sources. We take 418 duet videos to perform testing as illustrated in \cref{teaser fig}. 

~\\
\noindent \textbf{Metrics}
\quad
To measure the quality of separation \cite{mir_eval}, we adopt the widely-used mir$\_$eval library metrics: Signal-to-Distortion Ratio (SDR) measures both interference and artifacts, Signal-to-Interference Ratio (SIR) measures interference. Higher values indicate a better degree of separation. 

~\\
\noindent \textbf{Implementation Details}
\quad
We train our LAVSS framework with the implementation of PyTorch. 
We re-sample the audio at 11025Hz to get approximately 5.9s clip for each video. Then we perform STFT frame length of size 1022 and hop length of 256 \cite{sop,2.5D} to convert the time domain signal into 2D magnitude spectrogram of T, F = 256 after re-sampling to a log-frequency scale. We set the frame rate as 8fps and randomly select one frame per 5.9s video. 
We resize and crop the detected bounding boxes to $224\times224$ as the input of the ResNet-18 network. The MLP consists of two layers of 256, 512. All the attention modules are set of 8 heads and 2 decoder layers. In \cref{loss} the $\alpha$ and $\beta$ are set to 0.5 and 0.25, respectively. We apply Adam optimizer with $\beta_1=0.9$ and a weight decay of 1e-4. 
Since the MAVS and AVSS tasks are mutually related, we need to learn good initial models for AVSS. We start by pre-training on the MUSIC dataset to train the vision and sound separation network. Secondly, we introduce the IPD feature and co-learn the position network on FAIR-Play initialized with the pre-trained weights. The evaluation details are illustrated in the supplementary material.

\subsection{Audio-Visual Sound Separation}
\noindent \textbf{Comparison with State-of-the-Art}
\quad
To evaluate the performance of our LAVSS framework on audio-visual sound separation, we compare it to two baselines most related to binaural audio separation and generation: 2.5D Separation \cite{2.5D} and Sep-Stereo \cite{sep-stereo}, and recent state-of-the-art methods: SoP \cite{sop}, Co-separation \cite{co-sep}, and CCoL \cite{ccol}. 
\begin{figure*}[htbp]
\begin{minipage}[b]{1.0\linewidth}
  \centering
  \centerline{\includegraphics[width=14.3cm]{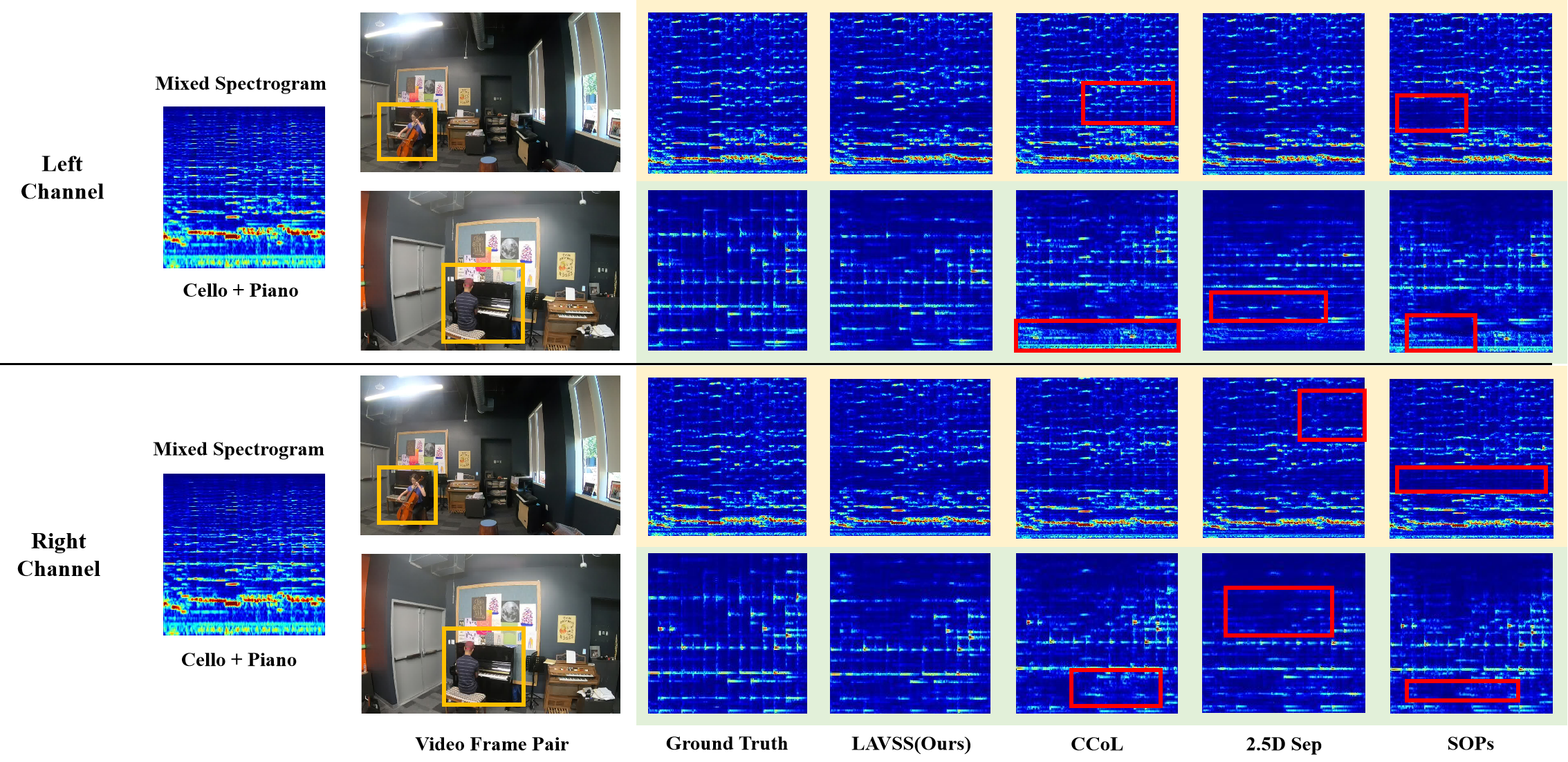}}
\end{minipage}
\caption{A set of solo separation results on FAIR-Play test set. Predicted spectrograms of SOTA methods and LAVSS are depicted for both channels. Red boxes illustrate the difference between the predicted spectrogram and the ground truth.}
\label{solospec}
\end{figure*}
\begin{figure*}[htbp]
\begin{minipage}[b]{1.0\linewidth}
  \centering
  \centerline{\includegraphics[width=13cm]{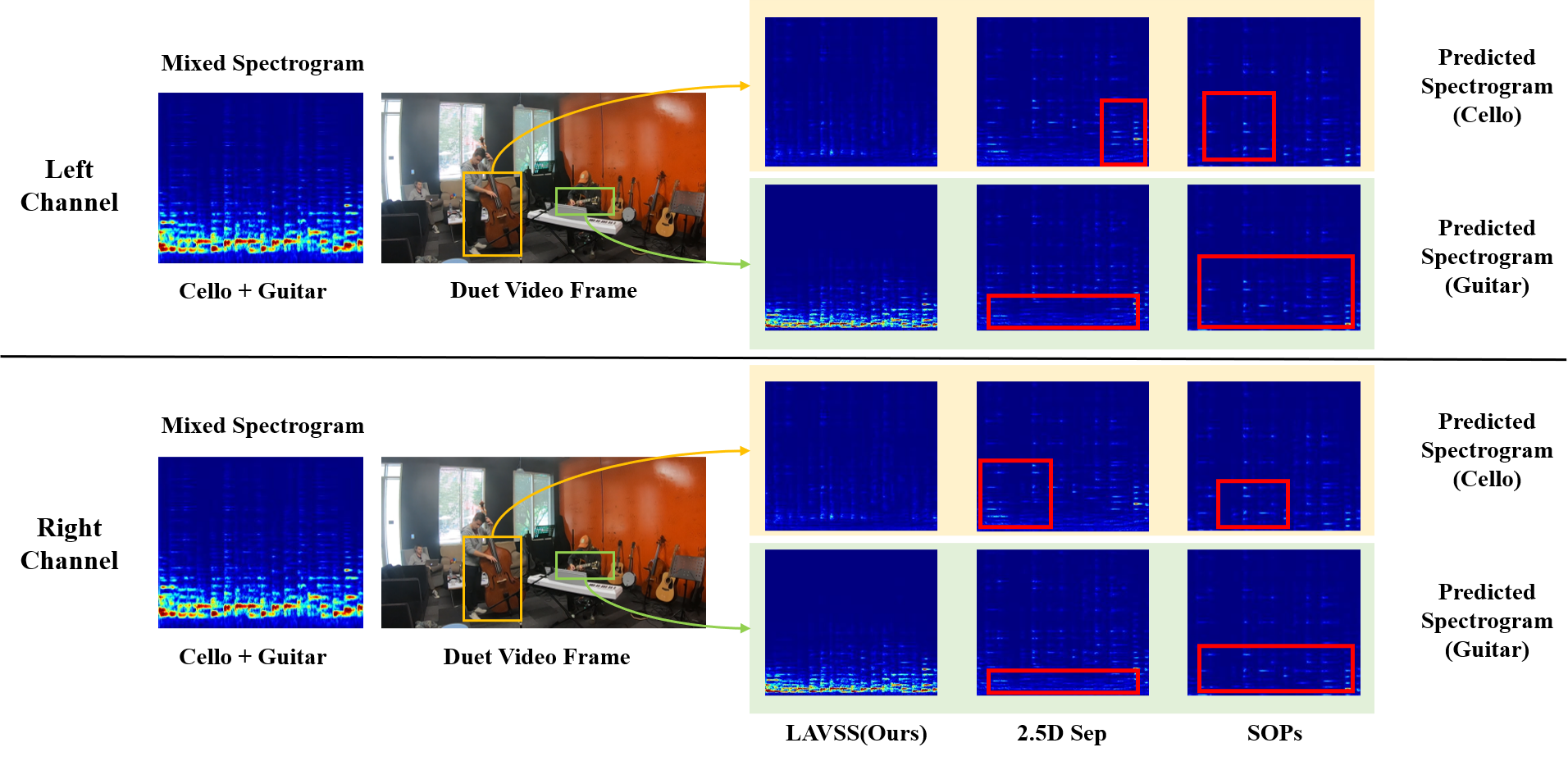}}
\end{minipage}
\caption{A set of duet separation results on FAIR-Play. Predicted spectrograms (cello and guitar) of SOTA methods and LAVSS are depicted for both channels. Red boxes indicate a comparison of separation ability between LAVSS and benchmarks.}
\label{duetspec}
\end{figure*}
\renewcommand{\arraystretch}{1.1} 
\begin{table}[b]
\centering
\resizebox{\linewidth}{!}{
\begin{tabular}{ccccccc}
\toprule
\multicolumn{1}{c}{\multirow{2}{*}{Method}}
&\multicolumn{2}{c}{Left Channel}
&\multicolumn{2}{c}{Right Channel}
&\multicolumn{2}{c}{Average}
\\
\cmidrule(lr){2-3}
\cmidrule(lr){4-5}
\cmidrule{6-7}
\multicolumn{1}{c}{} &{SDR$\uparrow$}&{SIR$\uparrow$}&{SDR$\uparrow$}&{SIR$\uparrow$}&{SDR$\uparrow$}&{SIR$\uparrow$}
\\
\midrule
\multicolumn{1}{c}{SoP\cite{sop}} & 3.98 & 7.03 & 3.96 & 6.99 & 3.97 & 7.01 \\
\multicolumn{1}{c}{2.5D\cite{2.5D}} & 4.44 & 8.20 & 4.47 & 8.26 & 4.45 & 8.23 \\
\multicolumn{1}{c}{Co-Sep\cite{co-sep}}  & 4.61 & 7.93 & 4.64 & 8.00 & 4.63 & 7.97  \\
\multicolumn{1}{c}{Sep-Stereo\cite{sep-stereo}} & 5.27 & 7.34  & 5.31 & 7.40 & 5.26 & 7.37 \\
\multicolumn{1}{c}{CCoL\cite{ccol}} & 5.05 & 8.89 & 5.17 & 9.02 & 5.11 & 8.96 \\
\midrule
\multicolumn{1}{c}{LAVSS (Ours)}& \textbf{5.89} & \textbf{10.08}  & \textbf{5.93} & \textbf{10.30} & \textbf{5.91} & \textbf{10.19} \\
\bottomrule
\end{tabular}}
\caption{ Comparisons of methods for source separation results on FAIR-Play test set. Higher is better for all metrics.}
\label{tabel1}
\end{table}
\renewcommand{\arraystretch}{1.2} 
\begin{table}[!htbp] 
\centering
\resizebox{\linewidth}{!}{
\begin{tabular}{ccccccc} 
\toprule
Models & Cello & Drum & Guitar & Harp & Piano & Trumpet  \\
\hline 
\rule{0pt}{8pt}
SoP & -2.12 & -1.88 & -2.69 & -1.77 & -2.35 & -1.78 \\ 
2.5D-sep & -0.93 & 0.86 & -2.00 & -1.49 & 0.35 & -2.55 \\
CCoL & -1.75 & -1.48 & 0.19 & -0.55 & -0.90 & -0.54  \\
Sep-Stereo & -1.32 & 0.34 & 1.68 & -0.41 & 1.38 & 0.73 \\ 
mix-gt & 0.58 & 0.25 & 0.67 & 0.47 & 1.17 & 1.51 \\
LAVSS(ours) & \textbf{1.53} & \textbf{2.67} & \textbf{3.72} & \textbf{1.83} & \textbf{3.13} & \textbf{3.12} \\
\bottomrule 
\end{tabular}}
\caption{The average separation results for both channels of the same instrument types from FAIR-Play in terms of SDR.}
\label{same type}
\end{table}

Note that five methods are evaluated for fair comparison on the FAIR-Play binaural dataset (including audio pre-processing) as ours. Since those methods are specialized in MAVS, we take the left and right channels into the network separately for training (after pre-training on 
 the MUSIC) and evaluation. The SDR and SIR quantitative analysis are illustrated in \cref{tabel1}. The results show that our LAVSS model outperforms its closest competitor, Sep-Stereo \cite{sep-stereo}, by an obvious superiority of 0.65 dB on SDR and 2.82 dB on SIR for binaural channels. Notably, our LAVSS boosts the SDR and SIR metrics by 0.80dB and 1.23dB compared to the most recent baseline CCoL \cite{ccol}. The above MAVS methods mainly utilize appearance-based visual information, which cannot generalize to AVSS. In contrast, our LAVSS simultaneously considers what and where the object is, thus demonstrating competence for the AVSS task.

\noindent \textbf{Separating Sources of the Same Type}
\quad
The source type is one of the critical factors affecting the performance of the separation. When two sounds have similar acoustic properties, separation becomes more complicated. In this case, the appearance features can not provide useful cues regarding similar images, while the location information guidance is particularly critical. Consequently, we select instruments of the same category from the FAIR-Play dataset and compare the separation performance of LAVSS with the MAVS methods. 
SoP \cite{sop} and  CCoL \cite{ccol} are mainly based on appearance guidance, 2.5D-sep \cite{2.5D}, and Sep-Stereo \cite{sep-stereo} are associated with binaural audio generation. 
Furthermore, we illustrate a comparison result called "mix-gt" to intuitively measure the mix spectrogram with the ground truth. 

\begin{figure}[bthp]
\begin{minipage}[b]{1.0\linewidth}
  \centering
  \centerline{  \includegraphics[width=6.5cm]{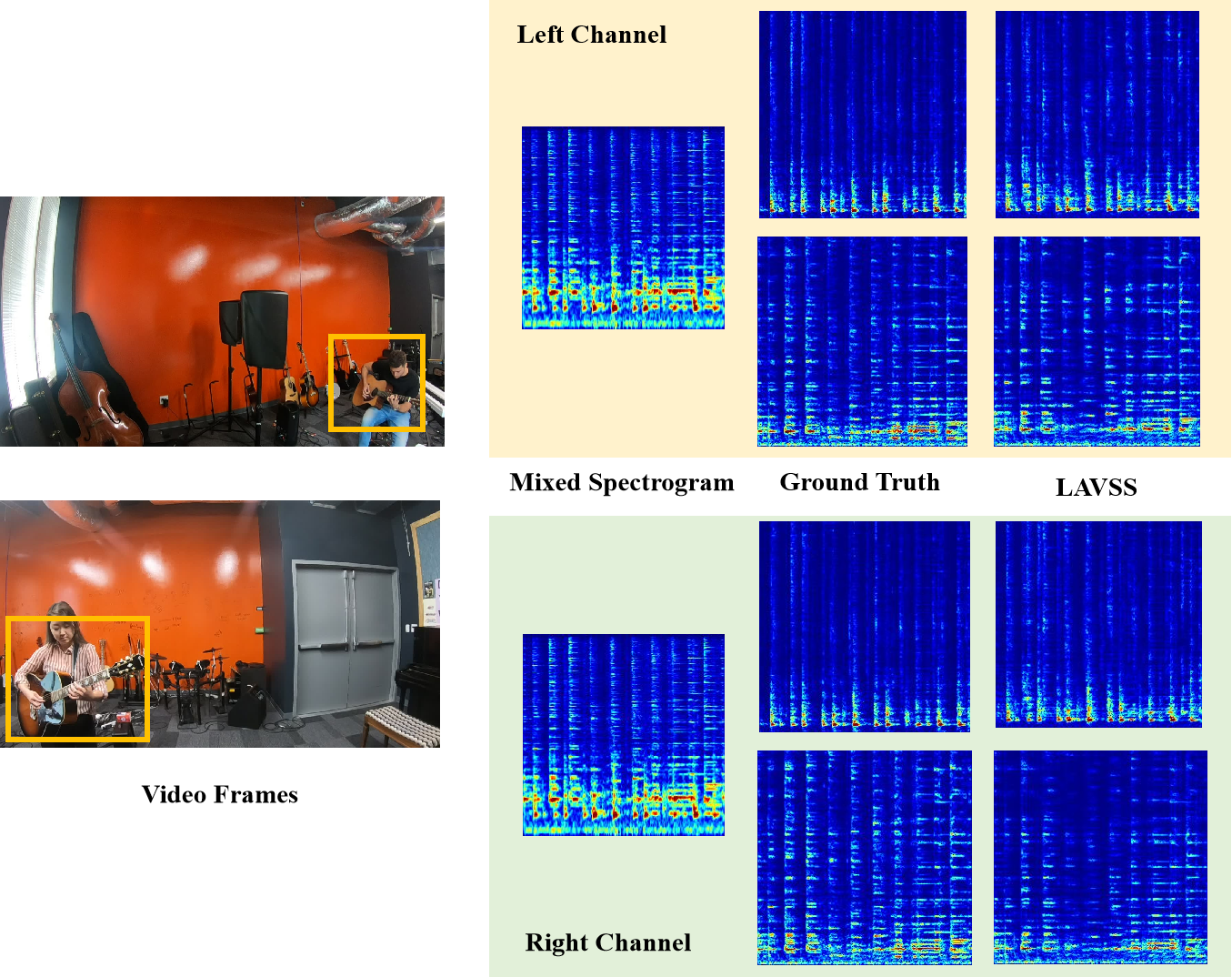}}
\end{minipage}
\caption{Illustration of the result for separating sound sources of the same type from the FAIR-Play dataset.}
\label{same_type_pic}
\end{figure}

Table~\ref{same type} demonstrates the averaged SDR results of both channels for cello, drum, guitar, harp, piano, and trumpet categories. The "mix-gt" baseline shows relatively better results in most cases, which indicates the challenges in monaural appearance-based models for spatial audio separation. Our method outperforms all MAVS baselines for all categories. The CCoL specifies the combinations of instruments selected for different types during training. The Sep-Stereo artificially rearranges the visual images and ignores the original location in the video frame. \cref{same_type_pic} shows a case of separating the sound mixture of the same type in different locations. Our method confirms that the relationship between object position and spatial phase cues brings significant improvement in separating similar sources.  

\subsection{Ablation Study and Performace Verification}
\noindent \textbf{Ablations of modality configurations}
\quad
We conduct ablation study to evaluate the effectiveness of IPD, position representation, and monaural transfer learning. 
We choose SoP and 2.5D-sep as baselines for verifying the versatility on benchmark applications. Note that the fusion strategy in \cref{attention} are applied for both baselines. \cref{tabel2} demonstrates the best scores when all ablation variants are applied, which confirms that the combined setup can be applied to any existing MAVS benchmarks to boost generalization ability.

One of the essential strategies we perform to strengthen AVSS is to explore position representation as a new modality for guidance. Rows 2 and 9 in \cref{tabel2} overwhelmingly point out the effectiveness of position encoding. Interestingly, we observe that the combination of the position feature and IPD are mutually beneficial since the network learns spatial location from both binaural audio and visual object. As a result, excavating the spatial properties of binaural audio brings 1.42dB and 2.03dB improvement in SDR and SIR, respectively.
We also explore the contribution of transfer pre-training on the external mono dataset. “*” denotes without fine-tuning on the FAIR-Play dataset. Rows 4-6 and 11-13 confirm that it definitely brings about 45$\%$ overall performance improvement on both metrics.

\vspace{-2mm}
\noindent \textbf{Ablations of multi-modal module design}
\quad
Ablation results of multi-scale audio fusion network design on FAIR-Play dataset are shown in \cref{attention_ablation}. “Tile-Concat” means the vanilla structure of replicating $F_{vp}$ to fit the audio feature $F_{a}$ at the bottleneck and performing concatenation through channel dimension. “w/o VP/AVP atten.” means removing the CMA block. Row 2 and 3 demonstrate that multi-modal fusion based on cross attention promotes stable and improved performance of sound source separation. “w/o multi” indicates that $F_{a}$ is only composed of tensor extracted by the last down-sample convolutional layer. Notably, multi-scale feature extraction increases the discrimination of audio representations and yields good results.  
\renewcommand{\arraystretch}{0.8} 
\begin{table}[t]
\centering
\resizebox{\linewidth}{!}{
\begin{tabular}{cccccccc}
\toprule
\multicolumn{1}{c}{\multirow{2}{*}{\makecell[c] {Baseline \\ Model}}}
&\multicolumn{1}{c}{\multirow{2}{*}{\makecell[c]{Position \\ Guidance}}}
&\multicolumn{1}{c}{\multirow{2}{*}{IPD}}
&\multicolumn{1}{c}{\multirow{2}{*}{\makecell[c]{Monaural \\ Pre-train}}}
&\multicolumn{2}{c}{Left Channel}
&\multicolumn{2}{c}{Right Channel}
\\
\cmidrule(lr){5-6}
\cmidrule{7-8}
\multicolumn{4}{c}{} &{SDR$\uparrow$}&{SIR$\uparrow$}&{SDR$\uparrow$}&{SIR$\uparrow$}
\\
\midrule
\multirow{8}{*}{SoP}
& \XSolidBrush & \XSolidBrush & \XSolidBrush & 3.34 & 6.45 & 3.29 & 6.42 \\
& \Checkmark & \XSolidBrush & \XSolidBrush & 4.00 & 7.31 & 4.02 & 7.27\\
& \Checkmark & \Checkmark & \XSolidBrush & 4.32 & 7.90 & 4.38 & 7.86\\
& \XSolidBrush & \XSolidBrush & \Checkmark * & 4.22 & 7.84 & 4.23 & 7.88\\
& \XSolidBrush & \XSolidBrush & \Checkmark  & 4.79 & 8.36 & 4.82 & 8.39\\
& \XSolidBrush & \Checkmark & \Checkmark & 5.14 & 8.57 & 5.15 & 8.55  \\
& \Checkmark & \Checkmark & \Checkmark & \textbf{5.32} & \textbf{8.71} & \textbf{5.36} & \textbf{8.73} \\

\midrule
\multirow{8}{*}{2.5D-sep}
& \XSolidBrush & \XSolidBrush & \XSolidBrush & 3.85 & 7.24 & 3.73 & 7.44 \\
& \Checkmark & \XSolidBrush & \XSolidBrush & 4.90 & 8.38 & 4.82 & 8.48 \\
& \Checkmark & \Checkmark & \XSolidBrush & 5.27 & 9.27 & 5.25 & 9.28\\
& \XSolidBrush & \XSolidBrush & \Checkmark * & 4.67 & 8.02 & 4.70 & 8.03\\
& \XSolidBrush & \XSolidBrush & \Checkmark & 5.03 & 8.56 & 5.08 & 8.59\\
& \XSolidBrush & \Checkmark & \Checkmark & 5.53 & 9.14 & 5.59 & 9.18  \\
& \Checkmark & \Checkmark & \Checkmark & \textbf{5.89} & \textbf{10.08}  & \textbf{5.93} & \textbf{10.30} \\
\bottomrule
\end{tabular}}
\caption{Ablation study of two benchmarks on FAIR-Play test set.}
\label{tabel2}
\end{table}
\renewcommand{\arraystretch}{0.9} 
\begin{table}[t]
\centering
\resizebox{\linewidth}{!}{
\begin{tabular}{ccccccc}
\toprule
\multicolumn{1}{c}{\multirow{2}{*}{Architecture}}
&\multicolumn{2}{c}{Left Channel}
&\multicolumn{2}{c}{Right Channel}
&\multicolumn{2}{c}{Average}
\\
\cmidrule(lr){2-3}
\cmidrule(lr){4-5}
\cmidrule{6-7}
\multicolumn{1}{c}{} &{SDR$\uparrow$}&{SIR$\uparrow$}&{SDR$\uparrow$}&{SIR$\uparrow$}&{SDR$\uparrow$}&{SIR$\uparrow$}
\\
\midrule
\multicolumn{1}{c}{LAVSS (Ours)}& \textbf{5.89} & \textbf{10.08}  & \textbf{5.93} & \textbf{10.30} & \textbf{5.91} & \textbf{10.19} \\
\multicolumn{1}{c}{w/o VP atten.} & 5.27 & 9.34 & 5.16 & 9.37 & 5.22 & 9.36 \\
\multicolumn{1}{c}{w/o AVP atten.}  & 5.04 & 8.63 & 5.05 & 8.65 & 5.04 & 8.64  \\
\multicolumn{1}{c}{w/o multi.} & 4.90 & 8.38 & 4.82 & 8.48 & 4.86 & 8.43 \\
\multicolumn{1}{c}{Tile-Concat} & 4.83 & 8.41 & 4.81 & 8.38 & 4.82 & 8.40 \\
\bottomrule
\end{tabular}}
\caption{ Ablations on the design of multi-modal attention module.}
\label{attention_ablation}
\end{table}

\noindent \textbf{Qualitative evaluation}
\quad
Specifically, both solo and duet video separation performances are illustrated in \cref{solospec,duetspec}. 
Our separated spectrogram is distinctly and completely restored for both channels compared to SoP, 2.5D-sep, and CCoL. 
For the duet case, the separation results of MAVS hardly show any difference. 
More qualitative separation results are revealed in the supplementary material.

\section{Conclusion}
In this work, we present LAVSS, a novel location-guided audio-visual spatial audio separator. We break through the limitation of MAVS methods and put forward AVSS. Our network exploits the synchronization between phase attributes of spatial audio and position embeddings of objects. 
We leverage location representations of objects and perform fusion with the visual information to consistently guide AVSS. 
Furthermore, we demonstrate the correlation of monaural and binaural channels by pre-training on external mono dataset for network transfer learning, which outperforms SOTA methods on FAIR-Play.
Discussions and future works are provided in the supplementary material. 

~\\
\noindent \textbf{Acknowledgement} This work was partly supported by the Foundations for the Development of Strategic Emerging sIndustries of Shenzhen(Nos.JSGG20211108092812020\&CJGJZD202104080\\92804011).


{\small
\bibliographystyle{ieee_fullname}
\bibliography{egbib}
}

\end{document}